\documentclass[twocolumn]{revtex4}
\usepackage{mathbbold}
\usepackage{amsfonts}
\usepackage{amsmath}
\usepackage{amssymb}
\usepackage{charter}
\usepackage{graphicx}

\begin{document}

\title{Changing Fock matrix elements of two-mode squeezed vacuum state by
employing three conditional operations in one-sided lossy channel}
\author{Hao-liang Zhang$^{1}$, Hong-chun Yuan$^{2}$ and Xue-xiang Xu$%
^{1,\dag }$}
\affiliation{$^{1}$College of Physics and Communication Electronics, Jiangxi Normal
University, Nanchang 330022, China\\
$^{2}$College of Electrical and Information Engineering, Changzhou Institute
of Technology, Changzhou 213032, China\\
$^{\dag }$Corresponding author: xuxuexiang@jxnu.edu.cn }

\begin{abstract}
This paper focuses on changing Fock matrix elements of two-mode squeezed
vacuum state (TMSVS) by employing three conditional operations in one-sided
lossy channel. These three conditional operations include one-photon
replacement (OPR), one-photon substraction (OPS) and one-photon addition
(OPA). Indeed, three conditional quantum states have been generated from the
original TMSVS. Using the characteristic function (CF) representation of
quantum density operator, we derive the analytical expressions of their Fock
matrix elements, which depend on the interaction parameters, including the
squeezing parameter of the input TMSVS, the loss factor and the
transmissivity of the variable beam splitter. For convenience of discussion,
we only give the Fock matrices in the subspace span $\left\{ \left\vert
00\right\rangle ,\left\vert 01\right\rangle ,\left\vert 10\right\rangle
,\left\vert 02\right\rangle ,\left\vert 11\right\rangle ,\left\vert
20\right\rangle \right\} $ for these two-mode states. Obviously, the TMSVS
only has the populations in $\left\vert 00\right\rangle $\ and $\left\vert
11\right\rangle $ in such subspace. By comparing the generated states with
the TMSVS, we find that: (1) The generated state after OPR will remain the
populations in $\left\vert 00\right\rangle $\ and $\left\vert
11\right\rangle $, and add the populations in $\left\vert 10\right\rangle $
and $\left\vert 20\right\rangle $; (2) The generated state after OPS will
lost the populations in $\left\vert 00\right\rangle $\ and $\left\vert
11\right\rangle $, but add the populations in $\left\vert 10\right\rangle $
and $\left\vert 20\right\rangle $; (3) The generated state after OPA will
remain the population only in $\left\vert 11\right\rangle $ and add the
population in $\left\vert 01\right\rangle $.

\textbf{Keywords:} two-mode squeezed vacuum state, Fock matrix elements;
twin-Fock state, conditional measurement, beam splitter, characteristic
function
\end{abstract}

\maketitle

\section{Introduction}

Quantum state tomography, as a standard approach to characterize unknown
quantum state, is often to construct a density matrix from which some other
information can be inferred. People can retrieve desired information about a
quantum state by performing multiple tomographic measurements (also the
so-called projective measurements in different bases) \cite{a1}. The most
common measurement is the homodyne measurement. Through homodyne
measurement, one can obtain a large amount of data, which can be converted
into the state's density matrix and/or Wigner function by resorting to
mathematical methods, such as the inverse Randon transformation, the
pattern-function method and the likelihood maximization algorithm \cite{a2}.
In other words, the unknown quantum state can be reconstructed in the
representation of a density matrix, either in the quadrature basis or in the
photon-number (Fock) basis. Recently, complete information about excited
coherent states has been analyzed by the optical tomography by Almarashia
and co-worker \cite{a3}.

Theoretically, in order to find the information of quantum state, one choose
projection operator in photon-number (Fock) basis to obtain its Fock matrix
elements. Indeed, every quantum state can be expanded in the number state
space and has its unique Fock matrix elements. For a one-mode quantum state,
the density operator $\rho $ can be written as $\rho =\sum_{n,m=0}^{\infty
}p_{mn}\left\vert m\right\rangle \left\langle n\right\vert $ with $%
p_{mn}=\left\langle m\right\vert \rho \left\vert n\right\rangle $, in which
the Fock matrix elements $p_{mn}$ is corresponding to component $\left\vert
n\right\rangle \left\langle m\right\vert $\ in the space of density operator
$\rho $. It should be emphasized that the elements $p_{mn}$ represent
populations (diagonal term $n=m$, real number)\ or coherences (off-diagonal $%
n\neq m$, often complex number). For example, the familiar thermal state, as
a mixed state, has only the components $\left\vert n\right\rangle
\left\langle n\right\vert $ ($n=0,1,2,\cdots $) with population element $%
p_{nn}$. The coherent state, as a typical pure state, has the components $%
\left\vert n\right\rangle \left\langle m\right\vert $ ($n,m=0,1,2,\cdots $)
with elements $p_{nm}$ (population or coherence). The single-mode squeezed
vacuum state, also a pure state, has only the components $\left\vert
2n\right\rangle \left\langle 2m\right\vert $ ($n,m=0,1,2,\cdots $) with
elements $p_{2n,2m}$ \cite{1,2}, which often was used as the typical optical
field interacting with atomical system \cite{2a,2b}.

Similarly, for a two-mode case, the density operator $\rho _{ab}$ can be
written in the two-mode photon-number basis%
\begin{equation}
\rho _{ab}=\sum_{n_{1},m_{1},n_{2},m_{2}=0}^{\infty
}p_{n_{1}m_{1},n_{2}m_{2}}\left\vert n_{1}\right\rangle _{a}\left\vert
m_{1}\right\rangle _{b}\left\langle n_{2}\right\vert _{a}\left\langle
m_{2}\right\vert _{b},  \label{0}
\end{equation}%
with $p_{n_{1}m_{1},n_{2}m_{2}}=\left\langle n_{1}\right\vert
_{a}\left\langle m_{1}\right\vert _{b}\rho _{ab}\left\vert
n_{2}\right\rangle _{a}\left\vert m_{2}\right\rangle _{b}$, which shows Fock
matrix element $p_{n_{1}m_{1},n_{2}m_{2}}$\ corresponding to component $%
\left\vert n_{1}\right\rangle _{a}\left\vert m_{1}\right\rangle
_{b}\left\langle n_{2}\right\vert _{a}\left\langle m_{2}\right\vert _{b}$\
in the space of two-mode density operator $\rho _{ab}$ \cite{3,4}. It should
be noted that $\left\vert n\right\rangle _{a}\left\vert m\right\rangle _{b}$
is often written as $\left\vert nm\right\rangle $ in this paper. Because the
space of the two-mode density operator has infinite two-mode number bases
span $\left\{ \left\vert nm\right\rangle \right\} $ ($n,m=0,1,\cdots ,\infty
$), we only study the Fock matrix elements in the subspace span $\left\{
\left\vert 00\right\rangle ,\left\vert 01\right\rangle ,\left\vert
10\right\rangle ,\left\vert 02\right\rangle ,\left\vert 11\right\rangle
,\left\vert 20\right\rangle \right\} $ for convenience of discussion, whose
corresponding Fock matrix can also be expressed in Table I. \textbf{\ }As we
all know,\ the two-mode squeezed vacuum state (TMSVS) is the primary
entangled resource in continuous-variable system, which can be used to
implement many quantum protocols, including continuous version of
teleportation and quantum key distribution. In addition, as a maximum
entangled state, the TMSVS has also been chosen as the typical optical field
interacting with atomical system \cite{4a}. Indeed, as a typical two-mode
quantum state, the TMSVS has only the twin-Fock components with the form $%
\left\vert nn\right\rangle \left\langle mm\right\vert $ ($n,m=0,1,2,\cdots $%
). Needless to say, it is impossible to have non-twin-Fock components for\
the TMSVS. Therefore, it is urgent for us to think that whether we have some
ways to change the Fock matrix elements for the TMSVS by using quantum
operations. This is the key aim of our paper.
\begin{table}[h]
\caption{Fock matrix elements of two-mode density operator $\protect\rho $
in subspace span $\left\{ \left\vert 00\right\rangle ,\left\vert
01\right\rangle ,\left\vert 10\right\rangle ,\left\vert 02\right\rangle
,\left\vert 11\right\rangle ,\left\vert 20\right\rangle ,\cdots \right\} $}
\begin{center}
\begin{tabular}{cccccccc}
\hline\hline
$\rho $ & $\left\vert 00\right\rangle $ & $\left\vert 01\right\rangle $ & $%
\left\vert 10\right\rangle $ & $\left\vert 02\right\rangle $ & $\left\vert
11\right\rangle $ & $\left\vert 20\right\rangle $ & $\cdots $ \\ \hline\hline
$\left\langle 00\right\vert $ & $p_{00,00}$ & $p_{00,01}$ & $p_{00,10}$ & $%
p_{00,02}$ & $p_{00,11}$ & $p_{00,20}$ & $\cdots $ \\
$\left\langle 01\right\vert $ & $p_{01,00}$ & $p_{01,01}$ & $p_{01,10}$ & $%
p_{01,02}$ & $p_{01,11}$ & $p_{01,20}$ & $\cdots $ \\
$\left\langle 10\right\vert $ & $p_{10,00}$ & $p_{10,01}$ & $p_{10,10}$ & $%
p_{10,02}$ & $p_{10,11}$ & $p_{10,20}$ & $\cdots $ \\
$\left\langle 02\right\vert $ & $p_{02,00}$ & $p_{02,01}$ & $p_{02,10}$ & $%
p_{02,02}$ & $p_{02,11}$ & $p_{02,20}$ & $\cdots $ \\
$\left\langle 11\right\vert $ & $p_{11,00}$ & $p_{11,01}$ & $p_{11,10}$ & $%
p_{11,02}$ & $p_{11,11}$ & $p_{11,20}$ & $\cdots $ \\
$\left\langle 20\right\vert $ & $p_{20,00}$ & $p_{20,01}$ & $p_{20,10}$ & $%
p_{20,02}$ & $p_{20,11}$ & $p_{20,20}$ & $\cdots $ \\
$\vdots $ & $\vdots $ & $\vdots $ & $\vdots $ & $\vdots $ & $\vdots $ & $%
\vdots $ & $\ddots $ \\ \hline\hline
\end{tabular}%
\end{center}
\end{table}

In recent years, some conditional operations, such as photon replacement,
photon subtraction and photon addition, have attracted extensive attention
of researchers. In fact, these schemes are related to conditional
measurements, which are fruitful methods for quantum-state manipulation and
engineering \cite{5}. Many nonclassical states have been generated by
conditional measurements theoretically or experimentally. In general, two
quantum states in the two output ports of the lossless beam splitter (BS)
are quantum-mechanically correlated with each other. If appropriate
measurement is employed in one of the output ports, then conditional quantum
state is generated in the other output port \cite{6,7}. In some conditional
measurement schemes, new output state $\rho _{out}$ is generated from the
input state $\rho _{in}$ in the main channel and the difference happen in
the ancillary\ channel. For example, photon-replacement scheme has the
feature that $m$-photon Fock state is input and the same $m$-photon Fock
state is measured in ancillary\ channel. This strategy is also called as
\textquotedblleft quantum-optical catalysis\textquotedblright\ \cite{8,9,10}%
. The photon-subtraction scheme has the feature that $m$-photon Fock state
is input and the bigger $n$-photon Fock state is measured in ancillary\
channel \cite{11,12,13}. The photon-addition scheme has the feature that $m$%
-photon Fock state is input and the smaller $n$-photon Fock state is
measured in ancillary\ channel \cite{14,15}. These non-Gaussian conditional
operations have proven advantageous in many scenarios such as entanglement
enhancement \cite{16,17} and teleportation improvement \cite{18,19}. On the
other hand, the loss accompanied by conditional operations is unavoidable,
which must be considered in realistic situation. Of course, losses may in
principle be overcome by some quantum techniques \cite{20,21,22,23}.

Combining with the above ideas and approaches, we aim to change the Fock
matrix elements of the TMSVS by employing conditional operations and
considering the loss. The Fock matrix elements before and after operations
are compared. Analytical and numerical results will be given in details. The
paper is organized as follows: In Sec.2, we make a brief review of the TMSVS
and introduce its Fock matrix elements. Here we raise the question of how
their elements will be changed. In Sec.3, we induce three quantum states
from the TMSVS, whose density operators are given. Sec.4 gives the
analytical expressions of Fock matrix elements for three generated states.
Numerical calculations about elements are made by choosing given interaction
parameters in Sec.5. Our conclusions are summarized in the last section.

\section{Two-mode squeezed vacuum state and its Fock matrix elements}

In this section, we make a brief review of the TMSVS and introduce its Fock
matrix elements. To begin with, we introduce the two-mode squeezing operator
\cite{24}
\begin{equation}
S\left( r\right) =\exp [r(a^{\dag }b^{\dag }-ab)],  \label{1.1}
\end{equation}%
where $a$\ and $b$\ are the annihilation operators for the two modes and $r$
is the real squeezing parameter. By operating $S\left( r\right) $\ on the
two-mode vacuum $\left\vert 00\right\rangle $, we can obtain the TMSVS%
\begin{equation}
S\left( r\right) \left\vert 00\right\rangle =\sqrt{1-\lambda ^{2}}\exp
\left( \lambda a^{\dag }b^{\dag }\right) \left\vert 00\right\rangle
\label{1.2}
\end{equation}%
with $\lambda =\tanh r$. Obviously, the TMSVS can be written in number basis
as follows%
\begin{equation}
S\left( r\right) \left\vert 00\right\rangle =\sum_{n=0}^{\infty
}c_{n}\left\vert nn\right\rangle =c_{0}\left\vert 00\right\rangle
+c_{1}\left\vert 11\right\rangle +\cdots  \label{1.2a}
\end{equation}%
with $c_{n}=\lambda ^{n}\sqrt{1-\lambda ^{2}}$. The obvious fact is that the
TMSVS is only the superposition of the twin Fock state, that is, $\left\vert
00\right\rangle $, $\left\vert 11\right\rangle $, $\cdots $ . It's
impossible for the TMSVS to contain non-twin-Fock state $\left\vert
nm\right\rangle $ with $n\neq m$, such as $\left\vert 01\right\rangle $, $%
\left\vert 10\right\rangle $, $\left\vert 02\right\rangle $, $\left\vert
12\right\rangle $, $\cdots $ .

Correspondingly, the density operator of the TMSVS can be expressed as%
\begin{equation}
\rho _{TMSVS}=S\left( r\right) \left\vert 00\right\rangle \left\langle
00\right\vert S^{\dag }\left( r\right) .  \label{1.3}
\end{equation}%
Using Eq.(\ref{1.2a}), it can be also expanded as%
\begin{eqnarray}
\rho _{TMSVS} &=&c_{0}c_{0}\left\vert 00\right\rangle \left\langle
00\right\vert +c_{0}c_{1}\left\vert 00\right\rangle \left\langle
11\right\vert +\cdots  \notag \\
&&+c_{1}c_{0}\left\vert 11\right\rangle \left\langle 00\right\vert
+c_{1}c_{1}\left\vert 11\right\rangle \left\langle 11\right\vert +\cdots
\notag \\
&&+\cdots ,  \label{1.4}
\end{eqnarray}%
where the density operator only contains the twin-Fock components.
Obviously, in our considered subspace, we find that the TMSVS has only
component $\left\vert 00\right\rangle \left\langle 00\right\vert $\ with
population probability $c_{0}c_{0}=1-\lambda ^{2}$\ and component $%
\left\vert 11\right\rangle \left\langle 11\right\vert $\ with population
probability $c_{1}c_{1}=\lambda ^{2}\left( 1-\lambda ^{2}\right) $. Of
course, there are other two coherence terms (i.e., corresponding to
components $\left\vert 00\right\rangle \left\langle 11\right\vert $\ and $%
\left\vert 11\right\rangle \left\langle 00\right\vert $) as non-zero
elements in matrix, which is the coherence between $\left\vert
00\right\rangle $\ and $\left\vert 11\right\rangle $. Table II gives the
Fock matrix elements of the TMSVS.

Now another question arises: what can we do to make the TMSVS to contain
elements corresponding to non-twin-Fock components, such as $\left\vert
00\right\rangle \left\langle 01\right\vert $ and $\left\vert 01\right\rangle
\left\langle 10\right\vert $? This will be the focus of our following work.

\begin{table}[h]
\caption{Fock matrix elements of the TMSVS in subspace span $\left\{
\left\vert 00\right\rangle ,\left\vert 01\right\rangle ,\left\vert
10\right\rangle ,\left\vert 02\right\rangle ,\left\vert 11\right\rangle
,\left\vert 20\right\rangle ,\cdots \right\} $}
\begin{center}
\begin{tabular}{cccccccc}
\hline\hline
$\rho _{TMSVS}$ & $\left\vert 00\right\rangle $ & $\left\vert
01\right\rangle $ & $\left\vert 10\right\rangle $ & $\left\vert
02\right\rangle $ & $\left\vert 11\right\rangle $ & $\left\vert
20\right\rangle $ & $\cdots $ \\ \hline\hline
$\left\langle 00\right\vert $ & $1-\lambda ^{2}$ & $0$ & $0$ & $0$ & $%
\lambda \left( 1-\lambda ^{2}\right) $ & $0$ & $\cdots $ \\
$\left\langle 01\right\vert $ & $0$ & $0$ & $0$ & $0$ & $0$ & $0$ & $\cdots $
\\
$\left\langle 10\right\vert $ & $0$ & $0$ & $0$ & $0$ & $0$ & $0$ & $\cdots $
\\
$\left\langle 02\right\vert $ & $0$ & $0$ & $0$ & $0$ & $0$ & $0$ & $\cdots $
\\
$\left\langle 11\right\vert $ & $\lambda \left( 1-\lambda ^{2}\right) $ & $0$
& $0$ & $0$ & $\lambda ^{2}\left( 1-\lambda ^{2}\right) $ & $0$ & $\cdots $
\\
$\left\langle 20\right\vert $ & $0$ & $0$ & $0$ & $0$ & $0$ & $0$ & $\cdots $
\\
$\vdots $ & $\vdots $ & $\vdots $ & $\vdots $ & $\vdots $ & $\vdots $ & $%
\vdots $ & $\ddots $ \\ \hline\hline
\end{tabular}%
\end{center}
\end{table}

\section{Three quantum states induced from the TMSVS}

As shown in Fig.1, we construct three conceptual schemes to generate three
new quantum states from the TMSVS. Using the TMSVS as the initial optical
field and operating three kinds of non-Gaussian conditional operations in
one mode (channel) which has the loss, we study and compare the Fock matrix
elements of the final optical fields. Using the characteristic function (CF)
representation, we obtain the normal forms (denoted by $\colon \colon $) for
every density operator. Of course, the success probabilities are also
obtained.

The propagation of optical field in every scheme includes two input-output
processes and three stages. The two processes include loss and non-Gaussian
operation, respectively. For the sake of convenience, we describe the
optical field in every stages with their corresponding density operators as
follows%
\begin{equation}
\rho _{ab}^{\left( I\right) }\Rightarrow \rho _{ab}^{\left( II\right)
}\Rightarrow \rho _{ab}^{\left( III\right) }.  \label{2.1}
\end{equation}%
Next we shall give the exact expressions of the density operator in every
stages.
\begin{figure}[tbp]
\label{Fig1} \centering\includegraphics[width=0.6\columnwidth]{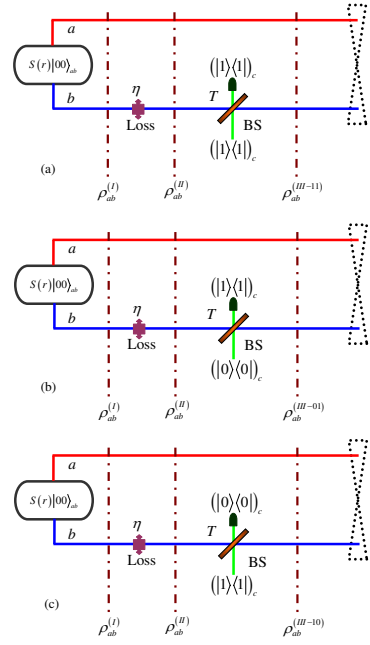}
\caption{(Colour online) Three conceptual quantum schemes to obtain quantum
states from the TMSV. The operations include (a) OPR, (b) OPS and (c) OPA,
where the loss is also considered. Here $\protect\eta $ is the loss factor
and $T$\ is transmissivity of the BS.}
\end{figure}

\textbf{Stage 1:} The corresponding density operator in stage 1 is $\rho
_{ab}^{\left( I\right) }=\rho _{TMSVS}$, where the initial optical field
under consideration is the TMSVS $S\left( r\right) \left\vert
00\right\rangle $. By using the Weyl expansion of the density operator, $%
\rho _{ab}^{\left( I\right) }$\ can also be expressed furtherly in the CF
representation as follows%
\begin{equation}
\rho _{ab}^{\left( I\right) }=\int \frac{d^{2}\alpha d^{2}\beta }{\pi ^{2}}%
\chi _{ab}^{\left( I\right) }\left( \alpha ,\beta \right) D_{a}\left(
-\alpha \right) D_{b}\left( -\beta \right) ,  \label{2.2}
\end{equation}%
where
\begin{eqnarray}
\chi _{ab}^{\left( I\right) }\left( \alpha ,\beta \right) &=&\mathrm{Tr}%
(\rho _{ab}^{\left( I\right) }D_{a}\left( \alpha \right) D_{b}\left( \beta
\right) )  \notag \\
&=&e^{-\frac{(1+\lambda ^{2})(\left\vert \alpha \right\vert ^{2}+\left\vert
\beta \right\vert ^{2})}{2(1-\lambda ^{2})}+\frac{\lambda (\allowbreak
\alpha ^{\ast }\beta ^{\ast }+\alpha \beta )}{1-\lambda ^{2}}}  \label{2.3}
\end{eqnarray}%
is just the CF of the TMSVS and $D_{a}\left( \alpha \right) =\exp \left(
\alpha a^{\dag }-\alpha ^{\ast }a\right) $ and $D_{b}\left( \beta \right)
=\exp \left( \beta b^{\dag }-\beta ^{\ast }b\right) $ are the displacement
operators, respectively.

\textbf{Stage 2:} In the second stage, we consider the loss only in channel
(mode) $b$, where the conditional operation will be employed. Thus, the
density operator of the optical field can be expressed as%
\begin{eqnarray}
\rho _{ab}^{\left( II\right) } &=&\int \frac{d^{2}\alpha d^{2}\beta }{\pi
^{2}}e^{-\frac{1}{2}\eta \left\vert \beta \right\vert ^{2}}\chi
_{ab}^{\left( I\right) }(\alpha ,\sqrt{1-\eta }\beta )  \notag \\
&&\times D_{a}\left( -\alpha \right) D_{b}\left( -\beta \right) ,
\label{2.4}
\end{eqnarray}%
where $\eta \in \left[ 0,1\right] $ is the loss factor. Here one can use the
formula derived in Ref.\cite{25} or see Appendix A.

\textbf{Stage 3:} There are the differences for the three sub-figures of
Fig.1 in this stage. Three conditional operations (that is, one-photon
replacement (OPR), one-photon subtraction (OPS) and one-photon addition
(OPA)) are used in three schemes, respectively.

\textit{Case OPR:} The OPR can be embodied in the $c$ mode, where one-photon
Fock state $\left\vert 1\right\rangle $\ is input and one-photon Fock state $%
\left\vert 1\right\rangle $ is measured \cite{26}. After employing the OPR,
we obtain the first generating state
\begin{equation}
\rho _{ab}^{\left( III-11\right) }=\frac{_{c}\left\langle 1\right\vert
\{B[\rho _{ab}^{\left( II\right) }\otimes \left( \left\vert 1\right\rangle
\left\langle 1\right\vert \right) _{c}]B^{\dag }\}\left\vert 1\right\rangle
_{c}}{p_{d}^{\left( 11\right) }},  \label{2.5}
\end{equation}%
where $p_{d}^{\left( 11\right) }$ is the success probability and $B=\exp %
\left[ \theta \left( b^{\dag }c-bc^{\dag }\right) \right] $\ is the BS
operator satisfying the following transformation
\begin{eqnarray}
BbB^{\dag } &=&\sqrt{T}b+\sqrt{1-T}c,  \notag \\
BcB^{\dag } &=&-\sqrt{1-T}b+\sqrt{T}c,  \label{2.5a}
\end{eqnarray}%
with transmissivity $T=\cos ^{2}\theta $. After detailed calculation, we
obtain the density operator in the normal ordering form%
\begin{align}
\rho _{ab}^{\left( III-11\right) }& =\frac{1-\lambda ^{2}}{p_{d}^{\left(
11\right) }}\frac{d^{4}}{dh_{2}dh_{1}ds_{2}ds_{1}}  \notag \\
& \colon e^{-(\allowbreak 1-\lambda ^{2}\eta )a^{\dag }a+\lambda \sqrt{%
1-\eta }\allowbreak \sqrt{1-T}\left( h_{2}a^{\dag }+h_{1}a\right) }  \notag
\\
& \times e^{-b^{\dag }b-\sqrt{1-T}\left( s_{1}b^{\dag }+s_{2}b\right)
+\lambda \sqrt{1-\eta }\sqrt{T}\left( a^{\dag }b^{\dag }+ab\right) }  \notag
\\
& \times e^{\sqrt{T}h_{2}s_{1}+\sqrt{T}h_{1}s_{2}}\colon
|_{s_{1}=s_{2}=h_{1}=h_{2}=0},  \label{2.6}
\end{align}%
where%
\begin{equation}
p_{d}^{\left( 11\right) }=\Omega ^{3}\left( 1-\lambda ^{2}\right) \left(
T+\kappa _{1}\lambda ^{2}+\kappa _{2}\lambda ^{4}\right)  \label{2.7}
\end{equation}%
is the success probability (also the normalization factor), whose derivative
form can be found in the appendix B, with%
\begin{eqnarray}
\Omega ^{-1} &=&1-\lambda ^{2}\eta -T\lambda ^{2}\left( 1-\allowbreak \eta
\right) ,  \notag \\
\kappa _{1} &=&\left( 1+T^{2}\right) \left( 1-\eta \right) +2T\left( \eta
-2\right) ,  \notag \\
\kappa _{2} &=&T-\eta \left( 1-\eta \right) \left( 1-T\right) ^{2}.
\label{2.8}
\end{eqnarray}%
Obviously, if $\eta \rightarrow 0$ and $T\rightarrow 1$, then $\rho
_{ab}^{\left( III-11\right) }\rightarrow \rho _{ab}^{\left( I\right) }$.
That is, the first generated state can be reduced to the TMSVS in this limit
case.

\textit{Case OPS:} The OPS can be embodied in the $c$ mode, where vacuum
state $\left\vert 0\right\rangle $\ is input and one-photon Fock state $%
\left\vert 1\right\rangle $ is measured. After employing the OPS, we obtain
the second generating state
\begin{equation}
\rho _{ab}^{\left( III-01\right) }=\frac{_{c}\left\langle 1\right\vert
\{B[\rho _{ab}^{\left( II\right) }\otimes \left( \left\vert 0\right\rangle
\left\langle 0\right\vert \right) _{c}]B^{\dag }\}\left\vert 1\right\rangle
_{c}}{p_{d}^{\left( 01\right) }},  \label{2.9}
\end{equation}%
where $p_{d}^{\left( 01\right) }$ is the success probability. The density
operator in the normal ordering form can be expressed as
\begin{align}
\rho _{ab}^{\left( III-01\right) }& =\frac{\allowbreak 1-\lambda ^{2}}{%
p_{d}^{\left( 01\right) }}\frac{d^{2}}{dh_{2}dh_{1}}  \notag \\
& \colon e^{-(\allowbreak 1-\lambda ^{2}\eta )a^{\dag }a+\lambda \sqrt{%
1-\eta }\allowbreak \sqrt{1-T}\left( h_{2}a^{\dag }+h_{1}a\right) }  \notag
\\
& \times e^{-b^{\dag }b+\lambda \sqrt{1-\eta }\sqrt{T}\left( a^{\dag
}b^{\dag }+ab\right) }\colon |_{h_{1}=h_{2}=0}  \label{2.10}
\end{align}%
with the success probability%
\begin{equation}
p_{d}^{\left( 01\right) }=\Omega ^{2}\lambda ^{2}\left( 1-\lambda
^{2}\right) \left( 1-\eta \right) \left( 1-T\right) .  \label{2.10a}
\end{equation}

\textit{Case OPA:} The OPA can be embodied in the $c$ mode, where one-photon
Fock state $\left\vert 1\right\rangle $\ is input and vacuum state $%
\left\vert 0\right\rangle $ is measured. After employing the OPA, we obtain
the third generating state
\begin{equation}
\rho _{ab}^{\left( III-10\right) }=\frac{_{c}\left\langle 0\right\vert
\{B[\rho _{ab}^{\left( II\right) }\otimes \left( \left\vert 1\right\rangle
\left\langle 1\right\vert \right) _{c}]B^{\dag }\}\left\vert 0\right\rangle
_{c}}{p_{d}^{\left( 10\right) }},  \label{2.11}
\end{equation}%
where $p_{d}^{\left( 10\right) }$ is the success probability. The density
operator in the normal ordering form can be expressed as%
\begin{eqnarray}
\rho _{ab}^{\left( III-10\right) } &=&\frac{\allowbreak 1-\lambda ^{2}}{%
p_{d}^{\left( 10\right) }}\frac{d^{2}}{ds_{2}ds_{1}}  \notag \\
&&\colon e^{-(\allowbreak 1-\lambda ^{2}\eta )a^{\dag }a+\lambda \sqrt{%
1-\eta }\sqrt{T}\left( a^{\dag }b^{\dag }+ab\right) }  \notag \\
&&\times e^{-b^{\dag }b-\sqrt{1-T}\left( s_{1}b^{\dag }+s_{2}b\right)
}\colon |_{s_{1}=s_{2}=0}  \label{2.12}
\end{eqnarray}%
with the success probability
\begin{equation}
p_{d}^{\left( 10\right) }=\Omega ^{2}\left( 1-\lambda ^{2}\right) \left(
1-\lambda ^{2}\eta \right) \left( 1-T\right) .  \label{2.12a}
\end{equation}

Obviously, these new generating states can be adjusted by the interaction
parameters, including the squeezing parameter $r$ of the input TMSV, the
loss factor $\eta $, and the transmissivity $T$ of BS.

\section{Fock matrix elements for three generated states}

Next, we study the Fock matrix elements for three generated states, which
will be compared with that of the original TMSVS. Firstly, we give their
analytical expressions for the Fock matrix elements and then give the Fock
matrix. As example, we plot the Fock matrices for these states and the
population elements by choosing given interaction parameters.

\subsection{Analytical expressions}

After detailed derivation, we obtain the following analytical expressions of
the Fock matrix elements for every quantum states.

(1) For $\rho _{ab}^{\left( III-11\right) }$, we have%
\begin{align}
p_{n_{1}m_{1},n_{2}m_{2}}^{\left( 11\right) }& =\frac{\allowbreak 1-\lambda
^{2}}{p_{d}^{\left( 11\right) }\sqrt{n_{1}!n_{2}!m_{1}!m_{2}!}}  \notag \\
& \frac{d^{n_{1}+n_{2}+m_{1}+m_{2}}}{%
dg_{2}^{m_{2}}dg_{1}^{m_{1}}df_{2}^{n_{2}}df_{1}^{n_{1}}}\frac{d^{4}}{%
ds_{1}ds_{2}dh_{1}dh_{2}}  \notag \\
& e^{\lambda \sqrt{1-\eta }\allowbreak \lbrack \sqrt{1-T}\left(
f_{1}h_{2}+h_{1}f_{2}\right) +\sqrt{T}\left( f_{1}g_{1}+f_{2}g_{2}\right) ]}
\notag \\
& \times e^{\sqrt{T}\left( h_{2}s_{1}+h_{1}s_{2}\right) -\sqrt{1-T}\left(
g_{1}s_{1}+g_{2}s_{2}\right) }  \notag \\
& \times e^{^{\lambda ^{2}\eta f_{1}f_{2}}\allowbreak
}|_{f_{1}=f_{2}=g_{1}=g_{2}=h_{1}=h_{2}=s_{1}=s_{2}=0}.  \label{3.1}
\end{align}

(2) For $\rho _{ab}^{\left( III-01\right) }$, we have%
\begin{align}
p_{n_{1}m_{1},n_{2}m_{2}}^{\left( 01\right) }& =\frac{\allowbreak 1-\lambda
^{2}}{p_{d}^{\left( 01\right) }\sqrt{n_{1}!n_{2}!m_{1}!m_{2}!}}  \notag \\
& \frac{d^{n_{1}+n_{2}+m_{1}+m_{2}}}{%
dg_{2}^{m_{2}}dg_{1}^{m_{1}}df_{2}^{n_{2}}df_{1}^{n_{1}}}\frac{d^{2}}{%
dh_{1}dh_{2}}  \notag \\
& e^{\lambda \sqrt{1-\eta }\allowbreak \lbrack \sqrt{1-T}\left(
f_{1}h_{2}+h_{1}f_{2}\right) +\sqrt{T}\left( f_{1}g_{1}+f_{2}g_{2}\right) ]}
\notag \\
& \times e^{\lambda ^{2}\eta f_{1}f_{2}\allowbreak
}|_{f_{1}=f_{2}=g_{1}=g_{2}=h_{1}=h_{2}=0}.  \label{3.2}
\end{align}

(3) For $\rho _{ab}^{\left( III-10\right) }$, we have%
\begin{align}
p_{n_{1}m_{1},n_{2}m_{2}}^{\left( 10\right) }& =\frac{\allowbreak 1-\lambda
^{2}}{p_{d}^{\left( 10\right) }\sqrt{n_{1}!n_{2}!m_{1}!m_{2}!}}  \notag \\
& \frac{d^{n_{1}+n_{2}+m_{1}+m_{2}}}{%
dg_{2}^{m_{2}}dg_{1}^{m_{1}}df_{2}^{n_{2}}df_{1}^{n_{1}}}\frac{d^{2}}{%
ds_{1}ds_{2}}  \notag \\
& e^{\lambda \sqrt{1-\eta }\allowbreak \sqrt{T}\left(
f_{1}g_{1}+f_{2}g_{2}\right) -\sqrt{1-T}\left( g_{1}s_{1}+g_{2}s_{2}\right) }
\notag \\
& \times e^{\lambda ^{2}\eta
f_{1}f_{2}}|_{f_{1}=f_{2}=g_{1}=g_{2}=s_{1}=s_{2}=0}.  \label{3.3}
\end{align}%
It should be noted that these expressions retain the derivative form because
of the complexity of components. With these expressions, we can calculate
the matrix elements in each case by mathematical software.

\subsection{Matrix forms and Numerical results}

In the considered subspace, we give the Fock matrix elements of state $\rho
_{ab}^{\left( III-11\right) }$ in Table III, where the non-zero elements
elements are
\begin{eqnarray}
p_{00,00}^{\left( 11\right) } &=&\frac{T\left( 1-\lambda ^{2}\right) }{%
p_{d}^{\left( 11\right) }},p_{10,10}^{\left( 11\right) }=\frac{\eta T\lambda
^{2}\left( 1-\lambda ^{2}\right) }{p_{d}^{\left( 11\right) }},  \notag \\
p_{00,11}^{\left( 11\right) } &=&p_{11,00}^{\left( 11\right) }=\frac{\sqrt{%
T\left( 1-\eta \right) }\lambda \left( 2T-1\right) \left( 1-\lambda
^{2}\right) }{p_{d}^{\left( 11\right) }},  \notag \\
p_{11,11}^{\left( 11\right) } &=&\frac{\left( 1-\eta \right) \left(
2T-1\right) ^{2}\lambda ^{2}\left( 1-\lambda ^{2}\right) }{p_{d}^{\left(
11\right) }},  \notag \\
p_{20,20}^{\left( 11\right) } &=&\frac{T\eta ^{2}\lambda ^{4}\left(
1-\lambda ^{2}\right) }{p_{d}^{\left( 11\right) }}.  \label{3.4}
\end{eqnarray}

\begin{table}[h]
\caption{Fock matrix elements of state $\protect\rho _{ab}^{\left(
III-11\right) }$ in subspace span $\left\{ \left\vert 00\right\rangle
,\left\vert 01\right\rangle ,\left\vert 10\right\rangle ,\left\vert
02\right\rangle ,\left\vert 11\right\rangle ,\left\vert 20\right\rangle
,\cdots \right\} $}
\begin{center}
\begin{tabular}{cccccccc}
\hline\hline
$\rho _{ab}^{\left( III-11\right) }$ & $\left\vert 00\right\rangle $ & $%
\left\vert 01\right\rangle $ & $\left\vert 10\right\rangle $ & $\left\vert
02\right\rangle $ & $\left\vert 11\right\rangle $ & $\left\vert
20\right\rangle $ & $\cdots $ \\ \hline\hline
$\left\langle 00\right\vert $ & $p_{00,00}^{\left( 11\right) }$ & $0$ & $0$
& $0$ & $p_{00,11}^{\left( 11\right) }$ & $0$ & $\cdots $ \\
$\left\langle 01\right\vert $ & $0$ & $0$ & $0$ & $0$ & $0$ & $0$ & $\cdots $
\\
$\left\langle 10\right\vert $ & $0$ & $0$ & $p_{10,10}^{\left( 11\right) }$
& $0$ & $0$ & $0$ & $\cdots $ \\
$\left\langle 02\right\vert $ & $0$ & $0$ & $0$ & $0$ & $0$ & $0$ & $\cdots $
\\
$\left\langle 11\right\vert $ & $p_{11,00}^{\left( 11\right) }$ & $0$ & $0$
& $0$ & $p_{11,11}^{\left( 11\right) }$ & $0$ & $\cdots $ \\
$\left\langle 20\right\vert $ & $0$ & $0$ & $0$ & $0$ & $0$ & $%
p_{20,20}^{\left( 11\right) }$ & $\cdots $ \\
$\vdots $ & $\vdots $ & $\vdots $ & $\vdots $ & $\vdots $ & $\vdots $ & $%
\vdots $ & $\ddots $ \\ \hline\hline
\end{tabular}%
\end{center}
\end{table}

Noticing that the coherence elements ($p_{00,11}^{\left( 11\right) }$\ or $%
p_{11,00}^{\left( 11\right) })$\ have the possibility of negative value if $%
\lambda \left( 2T-1\right) <0$. Comparing Table III of $\rho _{ab}^{\left(
III-11\right) }$\ and Table II for the TMSVS $\rho _{ab}^{\left( I\right) }$%
, we find that two new population elements (i.e., $p_{10,10}^{\left(
11\right) }$\ and $p_{20,20}^{\left( 11\right) }$) have been added, which
mean that there exist the populations in states $\left\vert 10\right\rangle $%
\ and $\left\vert 20\right\rangle $. Of course, the population probability
of $p_{00,00}^{\left( 11\right) }$\ and $p_{11,11}^{\left( 11\right) }$\
depends on the interaction parameter, which is also different from those of
the TMSVS.

Table IV gives the Fock matrix elements of state $\rho _{ab}^{\left(
III-01\right) }$, where the non-zero elements are%
\begin{eqnarray}
p_{10,10}^{\left( 01\right) } &=&\frac{\left( 1-T\right) \left( 1-\eta
\right) \lambda ^{2}\left( 1-\lambda ^{2}\right) }{p_{d}^{\left( 01\right) }}%
,  \notag \\
p_{20,20}^{\left( 01\right) } &=&2\eta \lambda ^{2}p_{10,10}^{\left(
01\right) }.  \label{3.5}
\end{eqnarray}%
It is surprising to find that the original elements of TMSVS are all
vanished but two new elements (i.e., $p_{10,10}^{\left( 11\right) }$\ and $%
p_{20,20}^{\left( 11\right) }$) have been added in this case. This point is
also like that OPR case.

\begin{table}[h]
\caption{Fock matrix elements of state $\protect\rho _{ab}^{\left(
III-01\right) }$ in subspace span $\left\{ \left\vert 00\right\rangle
,\left\vert 01\right\rangle ,\left\vert 10\right\rangle ,\left\vert
02\right\rangle ,\left\vert 11\right\rangle ,\left\vert 20\right\rangle
,\cdots \right\} $}
\begin{center}
\begin{tabular}{cccccccc}
\hline\hline
$\rho _{ab}^{\left( III-01\right) }$ & $\left\vert 00\right\rangle $ & $%
\left\vert 01\right\rangle $ & $\left\vert 10\right\rangle $ & $\left\vert
02\right\rangle $ & $\left\vert 11\right\rangle $ & $\left\vert
20\right\rangle $ & $\cdots $ \\ \hline\hline
$\left\langle 00\right\vert $ & $0$ & $0$ & $0$ & $0$ & $0$ & $0$ & $\cdots $
\\
$\left\langle 01\right\vert $ & $0$ & $0$ & $0$ & $0$ & $0$ & $0$ & $\cdots $
\\
$\left\langle 10\right\vert $ & $0$ & $0$ & $p_{10,10}^{\left( 01\right) }$
& $0$ & $0$ & $0$ & $\cdots $ \\
$\left\langle 02\right\vert $ & $0$ & $0$ & $0$ & $0$ & $0$ & $0$ & $\cdots $
\\
$\left\langle 11\right\vert $ & $0$ & $0$ & $0$ & $0$ & $0$ & $0$ & $\cdots $
\\
$\left\langle 20\right\vert $ & $0$ & $0$ & $0$ & $0$ & $0$ & $%
p_{20,20}^{\left( 01\right) }$ & $\cdots $ \\
$\vdots $ & $\vdots $ & $\vdots $ & $\vdots $ & $\vdots $ & $\vdots $ & $%
\vdots $ & $\ddots $ \\ \hline\hline
\end{tabular}%
\end{center}
\end{table}

Table V gives the Fock matrix elements of state $\rho _{ab}^{\left(
III-10\right) }$, where the non-zero elements are%
\begin{eqnarray}
p_{01,01}^{\left( 10\right) } &=&\frac{\left( 1-T\right) \left( 1-\lambda
^{2}\right) }{p_{d}^{\left( 10\right) }},  \notag \\
p_{11,11}^{\left( 10\right) } &=&\eta \lambda ^{2}p_{01,01}^{\left(
10\right) }.  \label{3.6}
\end{eqnarray}%
Here we find that only the population element of component $\left\vert
11\right\rangle \left\langle 11\right\vert $\ still exists and another
population element of component $\left\vert 01\right\rangle \left\langle
01\right\vert $ is added. This is also an interesting result.
\begin{table}[h]
\caption{Fock matrix elements of state $\protect\rho _{ab}^{\left(
III-10\right) }$ in subspace span $\left\{ \left\vert 00\right\rangle
,\left\vert 01\right\rangle ,\left\vert 10\right\rangle ,\left\vert
02\right\rangle ,\left\vert 11\right\rangle ,\left\vert 20\right\rangle
,\cdots \right\} $}
\begin{center}
\begin{tabular}{cccccccc}
\hline\hline
$\rho _{ab}^{\left( III-10\right) }$ & $\left\vert 00\right\rangle $ & $%
\left\vert 01\right\rangle $ & $\left\vert 10\right\rangle $ & $\left\vert
02\right\rangle $ & $\left\vert 11\right\rangle $ & $\left\vert
20\right\rangle $ & $\cdots $ \\ \hline\hline
$\left\langle 00\right\vert $ & $0$ & $0$ & $0$ & $0$ & $0$ & $0$ & $\cdots $
\\
$\left\langle 01\right\vert $ & $0$ & $p_{01,01}^{\left( 10\right) }$ & $0$
& $0$ & $0$ & $0$ & $\cdots $ \\
$\left\langle 10\right\vert $ & $0$ & $0$ & $0$ & $0$ & $0$ & $0$ & $\cdots $
\\
$\left\langle 02\right\vert $ & $0$ & $0$ & $0$ & $0$ & $0$ & $0$ & $\cdots $
\\
$\left\langle 11\right\vert $ & $0$ & $0$ & $0$ & $0$ & $p_{11,11}^{\left(
01\right) }$ & $0$ & $\cdots $ \\
$\left\langle 20\right\vert $ & $0$ & $0$ & $0$ & $0$ & $0$ & $0$ & $\cdots $
\\
$\vdots $ & $\vdots $ & $\vdots $ & $\vdots $ & $\vdots $ & $\vdots $ & $%
\vdots $ & $\ddots $ \\ \hline\hline
\end{tabular}%
\end{center}
\end{table}

By changing the interaction parameters, we can obtain the Fock matrix
elements by numerical simulation. Fig.2 shows the density matrices of the
TMSVS and other three quantum states with given parameters ($r=0.7$, $\eta
=0.2$, $T=0.7$). Here, we also give the corresponding numerical values for
the population elements. As we all know, the summarization of all population
elements are unity in principle. Because many other population elements
outside of our considered subspace are not given, we can't verify its unity
only in the considered subspace. Of course, one can calculate any population
element according to the corresponding expression. Furthermore, in order to
show the effect of the interaction parameters on these population elements,
we plot some population probabilities as the function of the loss factor $%
\eta $ or as a function of the transmissivity $T$ in Fig.3, Fig.4 and Fig.5,
where other parameters are fixed. One can see figures for details.
\begin{figure}[tbp]
\label{Fig2} \centering\includegraphics[width=0.8\columnwidth]{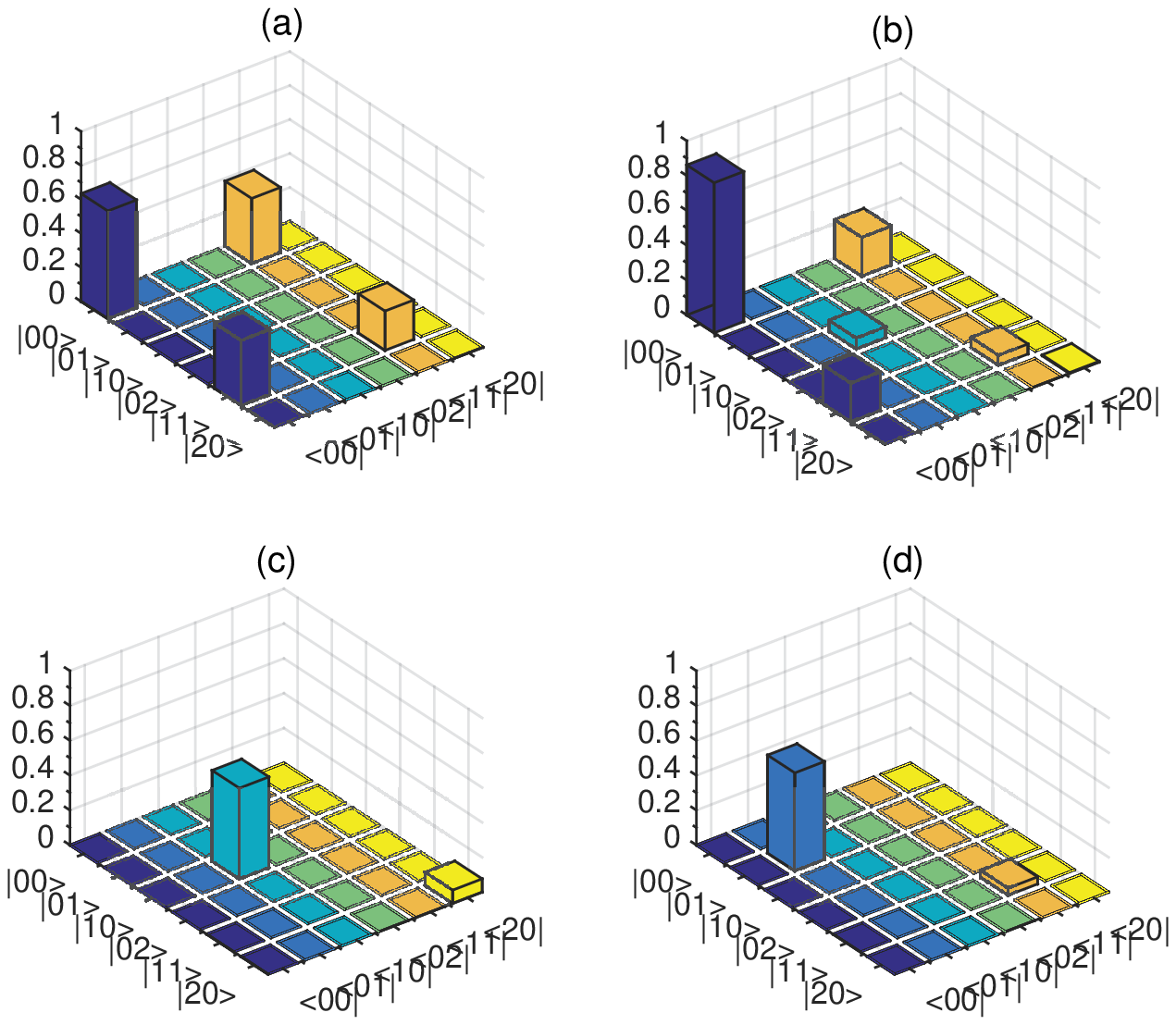}
\caption{Fock matrix elements of four states in our considered subspace. (a)
$\protect\rho _{ab}^{\left( I\right) }$; (b) $\protect\rho _{ab}^{\left(
III-11\right) }$; (c) $\protect\rho _{ab}^{\left( III-01\right) }$; (d) $%
\protect\rho _{ab}^{\left( III-10\right) }$, with $r=0.7$, $\protect\eta %
=0.2 $, $T=0.7$. In Fig.2 (a), $p_{00,00}=0.63474$, $p_{11,11}=0.231845$ and
$p_{00,11}=p_{11,00}=0.383616$; In Fig.2 (b), $p_{00,00}^{\left( 11\right)
}=0.861274$, $p_{10,10}^{\left( 11\right) }=0.0629178$, $p_{11,11}^{\left(
11\right) }=0.0575249$, $p_{20,20}^{\left( 11\right) }=0.00459628$, and $%
p_{00,11}^{\left( 11\right) }=p_{11,00}^{\left( 11\right) }=0.222586$; In
Fig.2 (c), $p_{10,10}^{\left( 01\right) }=0.521865$, $p_{20,20}^{\left(
01\right) }=0.0762466$; Fig.2 (d), $p_{01,01}^{\left( 10\right) }=0.562993$,
$p_{11,11}^{\left( 10\right) }=0.0411278$.}
\end{figure}
\begin{figure}[tbp]
\label{Fig3} \centering\includegraphics[width=0.8\columnwidth]{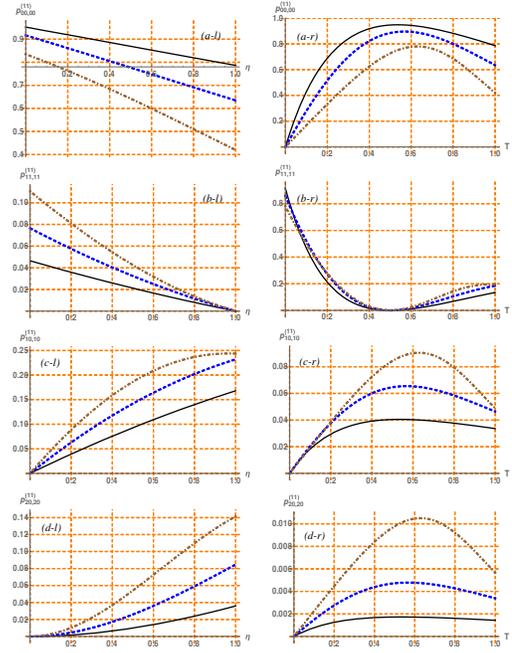}
\caption{Population elements (a) $p_{00,00}^{\left( 11\right) }$, (b) $%
p_{11,11}^{\left( 11\right) }$, (c) $p_{10,10}^{\left( 11\right) }$ and (d) $%
p_{20,20}^{\left( 11\right) }$ in state $\protect\rho _{ab}^{\left(
III-11\right) }$ as a function of $\protect\eta $\ with $T=0.7$ (left)\ or
as a function $T$ with $\protect\eta =0.2$\ (right). Here the black solid
line, blue dashed line and brown dotdashed line are corresponding to $r=0.5$%
, $r=0.7$\ and $r=1$, respectively.}
\end{figure}
\begin{figure}[tbp]
\label{Fig4} \centering\includegraphics[width=0.8\columnwidth]{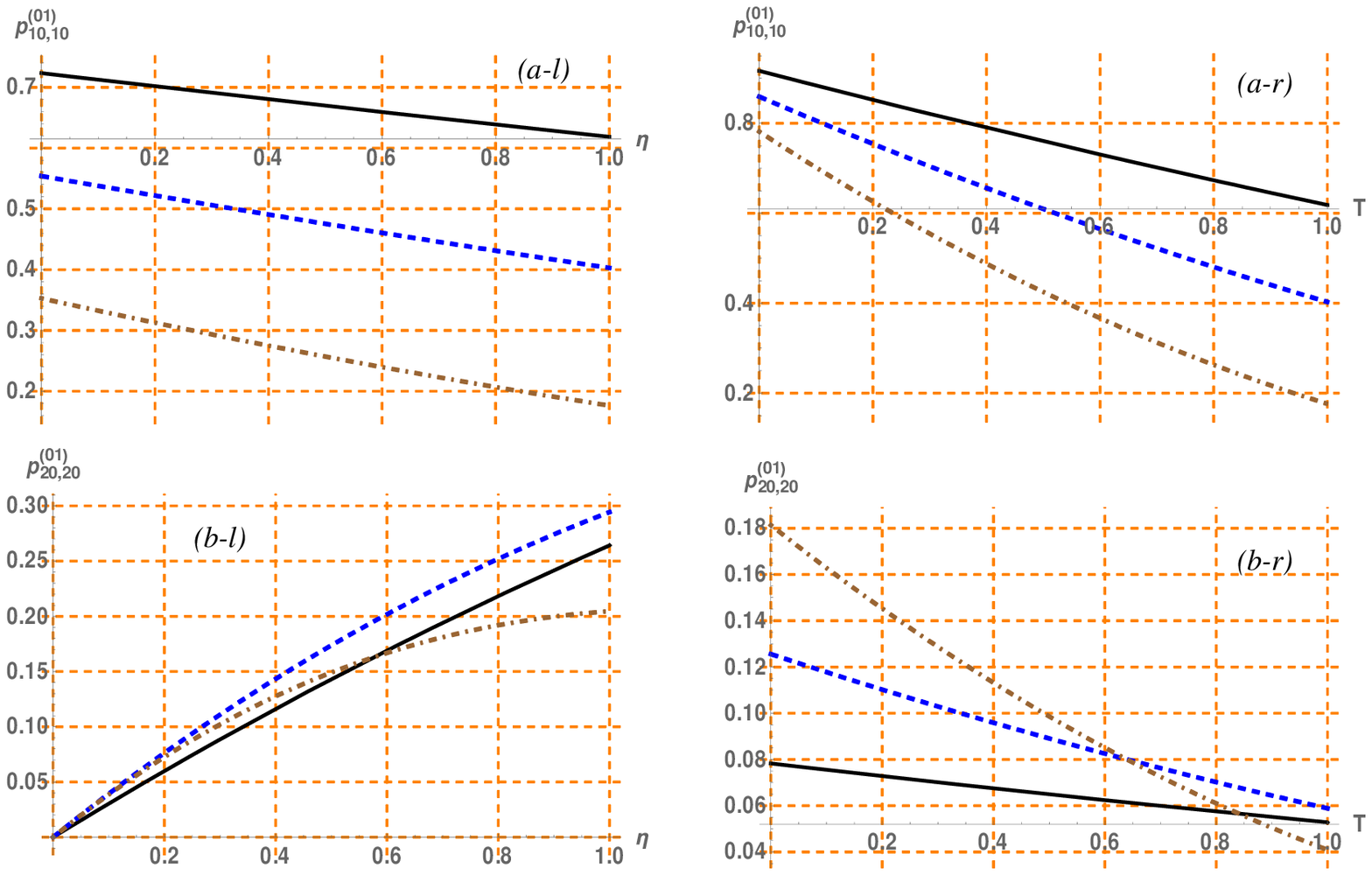}
\caption{Population elements (a) $p_{10,10}^{\left( 01\right) }$ and (b) $%
p_{20,20}^{\left( 01\right) }$\ in state $\protect\rho _{ab}^{\left(
III-01\right) }$ as a function of $\protect\eta $\ with $T=0.7$ (left)\ or
as a function $T$ with $\protect\eta =0.2$\ (right). Here the black solid
line, blue dashed line and brown dotdashed line are corresponding to $r=0.5$%
, $r=0.7$\ and $r=1$, respectively.}
\end{figure}
\begin{figure}[tbp]
\label{Fig5} \centering\includegraphics[width=0.8\columnwidth]{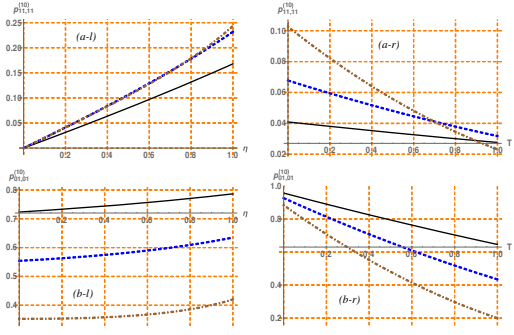}
\caption{Population elements (a) $p_{01,01}^{\left( 10\right) }$ and (b) $%
p_{11,11}^{\left( 10\right) }$\ in state $\protect\rho _{ab}^{\left(
III-10\right) }$ as a function of $\protect\eta $\ with $T=0.7$ (left)\ or
as a function $T$ with $\protect\eta =0.2$\ (right). Here the black solid
line, blue dashed line and brown dotdashed line are corresponding to $r=0.5$%
, $r=0.7$\ and $r=1$, respectively.}
\end{figure}

\section{Conclusions and discussions}

To summarize, we theoretically realized conditional operations to change the
Fock matrix elements of the TMSVS. These operations include OPR, OPS and
OPA. Everyone knows that the most prominent feature of the TMSVS is the
twin-number field, which leads to the population and coherence components
only from twin-Fock states. By employing three conditional operations, we
have prepared three entangled resources from the original TMSVS. We obtain
the analytical expressions of their Fock matrix elements and analyze the
change in elements. In the finite subspace span $\left\{ \left\vert
00\right\rangle ,\left\vert 01\right\rangle ,\left\vert 10\right\rangle
,\left\vert 02\right\rangle ,\left\vert 11\right\rangle ,\left\vert
20\right\rangle \right\} $, we compare their population elements, as shown
in Table VI. Compared with that the TMSVS only has the populations in $%
\left\vert 00\right\rangle $\ and $\left\vert 11\right\rangle $, we find
that (1) the populations in $\left\vert 10\right\rangle $ and $\left\vert
20\right\rangle $ have been added for the generated state after OPR; (2) the
populations in $\left\vert 00\right\rangle $\ and $\left\vert
11\right\rangle $ have been lost and the populations in $\left\vert
10\right\rangle $ and $\left\vert 20\right\rangle $ have been added for the
generated state after OPS; (3) the population only in $\left\vert
11\right\rangle $ has been remained and the population in $\left\vert
01\right\rangle $ has been added for the generated state after OPA. The
numerical results are shown in some figures.
\begin{table}[h]
\caption{Population elements in Fock matrix of these states in subspace span
$\left\{ \left\vert 00\right\rangle ,\left\vert 01\right\rangle ,\left\vert
10\right\rangle ,\left\vert 02\right\rangle ,\left\vert 11\right\rangle
,\left\vert 20\right\rangle \right\} $}
\begin{center}
\begin{tabular}{ccccccc}
\hline\hline
Population & $\left\vert 00\right\rangle $ & $\left\vert 01\right\rangle $ &
$\left\vert 10\right\rangle $ & $\left\vert 02\right\rangle $ & $\left\vert
11\right\rangle $ & $\left\vert 20\right\rangle $ \\ \hline\hline
$\rho _{ab}^{\left( I\right) }$ & $\surd $ &  &  &  & $\surd $ &  \\
$\rho _{ab}^{\left( III-11\right) }$ & $\surd $ &  & $\surd $ &  & $\surd $
& $\surd $ \\
$\rho _{ab}^{\left( III-01\right) }$ &  &  & $\surd $ &  &  & $\surd $ \\
$\rho _{ab}^{\left( III-10\right) }$ &  & $\surd $ &  &  & $\surd $ &  \\
\hline\hline
\end{tabular}%
\end{center}
\end{table}

In fact, there are a lot of related works. It is instructive to compare our
work with earlier related works. In order to improve figures of merit for a
quantum state, researchers often propose different quantum strategies, for
example, by applying the non-Gaussian operations (including photon addition,
photon subtraction, superposition of photon addition and subtraction) on the
original state. Of course, for a two-mode state such as the TMSVS, these
strategies may be applied to one or both modes. For example, Bartley and
Walmsley compare entanglement enhancement of the TMSVS after applying these
non-Gaussian operations \cite{11}. In addition, by means of quantum
catalysis \cite{26a} or quantum scissors \cite{26b}, the entanglement
properties of the TMSVS can also be enhanced in some parameter range. For
example, our group has two related works in recent years. One is by employed
local quantum catalysis on the TMSVS \cite{10}, another is by quantum
scissors \cite{26c} on the TMSVS. In fact, the OPR scheme in our work is
related to the idea of Ulanov and co-workers who proposed to distill the
TMSVS by using noiseless amplifiction \cite{21}. By the way, the OPS scheme
in our work is related to the idea of Kurochkin and co-workers who
demonstrated entanglement distillation by applying photon annihilation on
only one of the modes of the initial TMSVS \cite{26d}. Previous analyses of
these processes have adopted different figures of merit to compare each
protocol, for example teleportation fidelity, squeezing effect or
entanglement entropy. Unlike previous works, in this paper, we pay our
attention on the Fock matrix elements of the density operators for quantum
states under investigation.

Quantum technology protocols exploit the unique properties of quantum
systems to fulfill communication, computing and metrology tasks that are
impossible, inefficient or intractable for classical systems. Perhaps it is
because of the most prominent characteristic (eg. squeezing and
entanglement), the TMSVS have become the most commonly used entangled
resource of quantum technology \cite{27}. However, as our new states induced
from the TMSVS, they must have their own unique characteristic, which will
become new entangled resources to require the needs of quantum technologies.
The development of technologies allows promising real applications in
quantum information processing, such as quantum teleportation \cite{28},
quantum computation \cite{29} and quantum communication \cite{30}. It is
believed that our generating states will be good entangled resources for
future application. Our theoretical analyses will provide some information
for further application and stimulate the design of experimental tests.

\textbf{Appendix A: Loss formula in CF formalism}

About the detailed derivation of this loss formula, one can refer to our
previous work \cite{25}. The loss can be modelled in a BS formalism. The
input state $\rho _{in}$ can be expressed in the CF representation%
\begin{equation}
\rho _{in}=\int \frac{d^{2}\alpha }{\pi }\chi _{in}\left( \alpha \right)
D_{a}\left( -\alpha \right) ,  \tag{A.1}
\end{equation}%
where $D_{a}\left( \alpha \right) $ is the displacement operator in mode $a$%
, and $\chi _{in}\left( \alpha \right) =$Tr$\left( \rho _{in}D_{a}\left(
\alpha \right) \right) $ is the CF of the input state $\rho _{in}$. The
output state $\rho _{out}$ can be expressed as%
\begin{align}
\rho _{out}& =\int \frac{d^{2}\alpha }{\pi \left( 1-\eta \right) }e^{-\frac{%
\eta }{2\left( 1-\eta \right) }\left\vert \alpha \right\vert ^{2}}\chi
_{in}\left( \alpha \right)  \notag \\
& \times D_{a}(-\frac{\alpha }{\sqrt{1-\eta }}),  \tag{A.2}
\end{align}%
or
\begin{align}
\rho _{out}& =\int \frac{d^{2}\alpha }{\pi }e^{-\frac{1}{2}\eta \left\vert
\alpha \right\vert ^{2}}\chi _{in}(\sqrt{1-\eta }\alpha )  \notag \\
& \times D_{a}\left( -\alpha \right) ,  \tag{A.3}
\end{align}%
where $\eta $ is the loss factor. So, once the input CF $\chi _{in}\left(
\alpha \right) $ is known, one can obtain the output optical field after the
loss according to Eqs. (A.2) or (A.3).

$\allowbreak $\textbf{Appendix B: Success probability of generating states}

In order to ensure Tr$(\rho _{ab}^{\left( III\right) })=1$, we must
calculate the success probability for every scheme. These success
probabilities in derivative forms are given as follows, i.e.,%
\begin{align}
p_{d}^{\left( 11\right) }& =\Omega \left( 1-\lambda ^{2}\right) \frac{d^{4}}{%
dh_{2}dh_{1}ds_{2}ds_{1}}  \notag \\
& e^{\Omega \left( 1-T\right) [\lambda ^{2}\left( 1-\eta \right)
h_{1}h_{2}+(1-\lambda ^{2}\eta )s_{1}s_{2}]}  \notag \\
& \times e^{\Omega (1-\lambda ^{2})\sqrt{T}\left(
s_{1}h_{2}+h_{1}s_{2}\right) }|_{s_{1}=s_{2}=h_{1}=h_{2}=0},  \tag{B.1}
\end{align}%
and%
\begin{align}
p_{d}^{\left( 10\right) }& =\Omega \left( 1-\lambda ^{2}\right) \frac{d^{2}}{%
ds_{2}ds_{1}}  \notag \\
& e^{\Omega \left( 1-T\right) (1-\lambda ^{2}\eta
)s_{1}s_{2}}|_{s_{1}=s_{2}=h_{1}=h_{2}=0},  \tag{B.2}
\end{align}%
as well as%
\begin{align}
p_{d}^{\left( 01\right) }& =\Omega \left( 1-\lambda ^{2}\right) \frac{d^{2}}{%
dh_{2}dh_{1}}  \notag \\
& e^{\Omega \left( 1-T\right) \lambda ^{2}\left( 1-\eta \right)
h_{1}h_{2}}|_{s_{1}=s_{2}=h_{1}=h_{2}=0}.  \tag{B.3}
\end{align}

\begin{acknowledgments}
We acknowledge Bi-xuan Fan for helpful discussion. This project was
supported by the National Natural Science Foundation of China (No.11665013).
\end{acknowledgments}

\end{document}